\documentclass[preprint,12pt]{elsarticle}

\usepackage{amssymb}
\journal{Nuclear Physics A}

\begin{document}

\begin{frontmatter}



\title{Large mass dilepton production from jet-dilepton conversion in the quark-gluon plasma}


\author{Yong-Ping Fu, Yun-De Li}

\address{Department of Physics, Yunnan University, Kunming 650091, China}

\begin{abstract}
We calculate the production of large mass dileptons from the passage
of jets passing through the quark-gluon plasma. Using the
relativistic kinetic theory, we rigorously derive the production
rate for the jet-dilepton conversion in the hot medium. The
jet-dilepton conversion is compared with the thermal dilepton
emission and the Drell-Yan process. The contribution of the
jet-dilepton conversion is not prominent for all values of the
invariant mass $M$, and the Drell-Yan process is found to dominate
over the thermal dilepton emission and the jet-dilepton conversion
for $M>$2.5 GeV at RHIC. The jet-dilepton conversion is the dominant
source of large mass dileptons in the range of 4 GeV $<M<$10 GeV at
LHC.

\end{abstract}


\end{frontmatter}

% \linenumbers


\section{Introduction}

Finding the quark-gluon plasma (QGP) is one of the most important
goal in the studies of the relativistic heavy ion collisions. The
real and virtual photons are considered to be a useful probe for the
investigation of the evolution of the QGP due to their very long
mean free path. In the relativistic heavy ion collisions dileptons
are produced from various sources. These include the dileptons from
the Drell-Yan process \cite{DY1}, thermal dileptons from the QGP
\cite{th1,th2,th3,th4} and  the hadronic gas \cite{HG1,HG2,HG3},
dileptons from the jet-dilepton conversion in the hot medium
\cite{jetD1,jetD2}, and dileptons from the hadronic decays occurring
after the freeze-out \cite{FZ1,FZ2,FZ3}.

The measurement of the dilepton continuum at Relativistic Heavy Ion
Collider (RHIC) energies was performed by the PHENIX experiments for
Au+Au collisions at $\sqrt{s_{NN}}=$200 GeV \cite{DE1,DE2,DE3,DE4}.
The dilepton yield in the low mass range between 0.2 and 0.8 GeV is
enhanced by a factor of 2$\sim$3 compared to the expectation from
hadron decays. In fact, such phenomenon was also found at Super
Proton Synchrotron (SPS) \cite{DESPS1}, this dilepton excess at SPS
was successfully interpreted by the models of the dropping or
melting mass in a hot medium due to the chiral symmetry restoration,
but such modifying scenarios can not well explain the excess in the
low mass region at RHIC energies \cite{FZ1,ch1}.

In the intermediate mass region between $\phi$ and $J/\Psi$
resonances the dominant contribution arises from the correlated
decays of charm mesons \cite{jetD1,jetD2}. This region has been
suggested as a candidate to search for the thermal dilepton
emission, since its contribution could be comparable to that of
charm decays \cite{DE2}. The intermediate mass dilepton excess
observed by the NA50 and NA60 experiments has suggested that the
intermediate mass dileptons are partly produced from the QGP, and
not just charm decays \cite{NA1,NA2,NA3}.

In Refs. \cite{jetD1,jetD2} the authors have calculated the
production of large mass dileptons originating from the passage of
the jets passing through the QGP at leading order, and have
suggested that the jet-dilepton conversion as a new dilepton source
would confirm the occurrence of the jet-plasma interactions and the
existence of the QGP. The yield of large mass dileptons may be
enhanced by the new source of the dilepton production. However, the
numerical treatment used by Ref. \cite{jetD1} is not proper
\cite{jetD2}.

In the present work, we rigorously derive the production rate for
the jet-dilepton conversion by using the relativistic kinetic
theory. We compare the contribution of the jet-plasma interaction
with the thermal emission and the Drell-Yan process. Numerical
results indicate that the contribution of the jet-dilepton
conversion is not prominent at RHIC energies. The Drell-Yan process
is the dominant source of large mass dileptons for $M>$ 2.5 GeV at
RHIC. The jet-dilepton conversion starts playing an interesting role
at Large Hadron Collider (LHC) energies. The jet-plasma interactions
are found to dominate over the Drell-Yan process and the thermal
dilepton emission in the range of 4 GeV $<M<$10 GeV at LHC. By
comparing with the yield of the Drell-Yan process, we find that the
spectrum of the jet-dilepton conversion is reduced rapidly with the
invariant mass at RHIC and LHC energies.

This article is organized as follows. In Sec.2 we investigate the
production rate of the jet-dilepton conversion, the thermal dilepton
emission and the Drell-Yan process are also presented. The numerical
results and discussion are given in Sec.3. Finally, a summary is
presented in Sec.4.

\section{Formulation}
\subsection{Thermal dileptons and jet-dilepton conversion}

The quark jets crossing the hot and dense medium can produce large
mass dileptons by annihilation with the thermal antiquarks
($q_{jet}\bar{q}_{th}\rightarrow l^{+}l^{-}$ and
$q_{th}\bar{q}_{jet}\rightarrow l^{+}l^{-}$). By using the
relativistic kinetic theory, the production rate for the above
annihilation process can be written as
\begin{eqnarray}\label{eq1}
R_{jet-l^{+}l^{-}}=\int\frac{d^{3}p_{1}}{(2\pi)^{3}}\int\frac{d^{3}p_{2}}{(2\pi)^{3}}
f_{jet}(\emph{\textbf{p}}_{1})f_{th}(\emph{\textbf{p}}_{2})\sigma(M)v_{12}.
\end{eqnarray}
The cross section of the $q\bar{q}\rightarrow l^{+}l^{-}$
interaction is given by
\begin{eqnarray}\label{eq2}
\sigma(M)=\frac{4\pi}{3}\frac{\alpha^{2}}{M^{2}}N_{c}N_{s}^{2}\sum_{q}e_{q}^{2},
\end{eqnarray}
where the parameters $N_{c}$ and $N_{s}$ are the color number and
spin number, respectively. The relative velocity is
\begin{eqnarray}\label{eq3}
v_{12}=\frac{(p_{1}+p_{2})^{2}}{2E_{1}E_{2}}.
\end{eqnarray}
In the relativistic collisions, $|\emph{\textbf{p}}|\simeq E$, the
integration over
$d^{3}p=|\emph{\textbf{p}}|^{2}d|\emph{\textbf{p}}|d\Omega$ can be
done with the relatively simple result
\begin{eqnarray}\label{eq4}
\frac{dR_{jet-l^{+}l^{-}}}{dM^{2}}=\frac{\sigma(M)M^{2}}{2(2\pi)^{4}}
\int d|\emph{\textbf{p}}_{1}|\int d|\emph{\textbf{p}}_{2}|
f_{jet}(\emph{\textbf{p}}_{1})f_{th}(\emph{\textbf{p}}_{2}),
\end{eqnarray}
where the distribution of thermal partons is
$f_{th}(\emph{\textbf{p}})=\exp(-E/T)$. The limits of the
$d|\emph{\textbf{p}}_{2}|$ integration are given by
$\left[\infty,M^{2}/(4|\emph{\textbf{p}}_{1}|)\right]$ due to the
definition of the invariant mass
$M^{2}=(p_{1}+p_{2})^{2}=2|\emph{\textbf{p}}_{1}||\emph{\textbf{p}}_{2}|(1-\cos\theta_{\angle12})$.
Then we have
\begin{eqnarray}\label{eq5}
\frac{dR_{jet-l^{+}l^{-}}}{dM^{2}}=\frac{\sigma(M)M^{2}}{2(2\pi)^{4}}
\int d|\emph{\textbf{p}}_{1}|
f_{jet}(\emph{\textbf{p}}_{1})Te^{-\frac{M^{2}}{4|\emph{\textbf{p}}_{1}|T}}.
\end{eqnarray}
If the phase-space distribution for the quark jets
$f_{jet}(\emph{\textbf{p}})$ is replaced by the thermal distribution
$f_{th}(\emph{\textbf{p}})$ in Eq.(\ref{eq5}), one can obtain the
rate for producing thermal dileptons as \cite{th1}
\begin{eqnarray}\label{eq6}
\frac{dR_{th}}{dM^{2}}=\frac{\sigma(M)M^{3}}{2(2\pi)^{4}}
TK_{1}\left(\frac{M}{T}\right),
\end{eqnarray}
where the Bessel function is $K_{1}(z)=\sqrt{\pi/(2z)}e^{-z}$. If
the QGP is created in the relativistic heavy ion collisions, the
plasma may reach kinetic equilibrium quickly \cite{jetD1}. In 1+1
dimension Bjorken expansion, the system temperature evolves as
$T=T_{0}\left(\tau_{0}/\tau\right)^{1/3}$ \cite{Bjor1}, where
$\tau_{0}\sim 1/(3T_{0})$ is the initial time when the temperature
reaches $T_{0}$ \cite{Hy1,Hy2,Hy3}(see Table \ref{temperature}). In
the Bjorken model, the transverse density of nucleus is assumed to
be constant. Since a nucleus does have a transverse density profile,
the initial temperature of the system can be  assigned by the
transverse profile function as
$T(r,\tau_{0})=T_{0}[2(1-r^{2}/R^{2}_{\bot})]^{1/4}$ \cite{Hy1,Hy2}
while performing the space-time integration $d^{4}x=\tau d\tau rdr
d\eta d\phi$. The limits of the integration over the time $\tau$ are
$[\tau_{0},\tau_{c}]$ for the QGP phase and $[\tau_{c},\tau_{h}]$
for the mixed phase, such that
\begin{eqnarray}\label{eq7}
\int d\tau=\int_{\tau_{0}}^{\tau_{c}}
d\tau+\int_{\tau_{c}}^{\tau_{h}} d\tau f_{QGP}(\tau),
\end{eqnarray}
where $\tau_{c}=\tau_{0}(T_{0}/T_{c})^{3}$ is the critical time when
the QGP phase transfers into the mixed phase, and
$\tau_{h}=r_{d}\tau_{c}$ is the time when the mixed phase transfers
into the hadronic phase. The fraction of the QGP matter is
$f_{QGP}(\tau)=(r_{d}\tau_{c}/\tau-1)/(r_{d}-1)$, here
$r_{d}=g_{Q}/g_{H}$ is the ratio of the degrees of freedom in the
two phases, we have $g_{Q}=42.25$ for the three flavors of quarks
and $g_{H}=3$ for the hadronic gas of pions \cite{jetD2,Hy3}.

\begin{table}
\centering\caption{Initial conditions of the hydrodynamical
expansion \cite{jetD2}: initial time($\tau_{0}$), initial
temperature($T_{0}$) and critical temperature($T_{c}$).}
\label{temperature}%\vspace{-1mm}
\begin{tabular}{rcccccccc}
\hline Energy &$\tau_{0}(fm/c)$&$T_{0}$(MeV)&$T_{c}$(MeV)& \\
\hline
RHIC&$ 0.26 $&$ 370$&$ 160$&\\
LHC &$ 0.088$&$ 845$&$ 160$&\\
\hline
\end{tabular}
\end{table}

The phase-space distribution of the quark jets produced in the
relativistic heavy ion collisions is \cite{Hy1,Hy2}
\begin{eqnarray}\label{eq7}
f_{jet}(\emph{\textbf{p}})&=&\frac{(2\pi)^{3}}{g_{q}\pi
R_{\bot}^{2}\tau p_{T}\cosh
y}\frac{dN_{jet}}{d^{2}p_{T}dy}R(r)\delta(y-\eta)\nonumber\\[1mm]
&&\times\Theta(\tau-\tau_{i})\Theta(\tau_{max}-\tau)\Theta(R_{\bot}-r)
,
\end{eqnarray}
where $g_{q}=6$ is the spin and color degeneracy of the quarks (and
antiquarks), $R_{\bot}=1.2A^{1/3} fm$ is the transverse radius of
the system, $\eta$ is the space time rapidity,
$R(r)=2(1-r^{2}/R^{2}_{\bot})$ is the transverse profile function,
$\tau_{i}\sim1/p_{T}$ is the formation time of the quark or
antiquark jet, $\tau_{max}$ is smaller than the lifetime of the QGP
and the time taken by the jet produced at position
$\emph{\textbf{r}}$ to reach the surface of the plasma.

Jets crossing the hot and dense plasma will lose their energy.
Induced gluon bremsstrahlung, rather than elastic scattering of
partons, is the dominant contribution of the jet energy loss
\cite{Hy3,EL1,EL2}. Based on the AMY formulism \cite{EL2.1}, the
energy loss of the final state partons can be described as a
dependence of the final state parton spectrum $dN_{jet}/dE$ on time
\cite{Hy3}. The energy loss is scaled as the square of the distance
traveled through the hot medium \cite{EL3}. Jets travel only a short
distance through the plasma, and do not lose a significant amount of
energy. Quark jets lose energy at less than half the rate as gluon
jets, and the quark jets form the main fraction of jet events
\cite{Hy1}. The energy loss effect of jets before they convert into
dileptons is found to be small, just about 20$\%$ \cite{Hy2}.

\subsection{Jets production}

The cross section for producing jets in hadronic collisions
($A+B\rightarrow jets+X$) can be factored in the perturbative QCD
(pQCD) theory as \cite{Owen}
\begin{eqnarray}\label{eq8}
\frac{d\sigma_{jet}}{d^{2}p_{T}dy}=\sum_{a,b}\frac{1}{\pi}
\int_{x_{a}^{min}}^{1}dx_{a}G_{a/A}(x_{a},Q^{2})G_{b/B}(x_{b},Q^{2})
\frac{x_{a}x_{b}}{x_{a}-x_{1}}K_{jet}\frac{d\hat{\sigma}_{ab\rightarrow
cd}}{d\hat{t}},
\end{eqnarray}
where $x_{a}(x_{b})$ is the momentum fraction of the parton $a(b)$
of the nucleon $A(B)$. The momentum fractions with the rapidity $y$
are given by
\begin{eqnarray}\label{eq9}
x_{a}^{min}=\frac{x_{1}}{1-x_{2}},
\end{eqnarray}
\begin{eqnarray}\label{eq10}
x_{b}=\frac{x_{a}x_{2}}{x_{a}-x_{1}},
\end{eqnarray}
where the variables are $x_{1}=x_{T}e^{y}/2$, $x_{2}=x_{T}e^{-y}/2$,
$x_{T}=2p_{T}/\sqrt{s_{NN}}$. $p_{T}$ is the transverse momentum of
the final state partons, $\sqrt{s_{NN}}$ is the center of mass
energy of the colliding nucleons. The parton distribution for the
nucleus is given by
\begin{eqnarray}\label{eq11}
G_{a/A}(x_{a},Q^{2})=R^{a}_{A}(x_{a},Q^{2})\left[Zf_{a/p}(x_{a},Q^{2})+(A-Z)f_{a/n}(x_{a},Q^{2})\right]/A,
\end{eqnarray}
where $R^{a}_{A}(x_{a},Q^{2})$ is the nuclear modification of the
structure function \cite{shad1}, $Z$ is the number of protons, $A$
is the number of nucleons. The functions $f_{a/p}(x_{a},Q^{2})$ and
$f_{a/n}(x_{a},Q^{2})$ are the parton distributions of the proton
and neutron, respectively \cite{part1}. We choose $Q^{2}=p_{T}^{2}$.
$d\hat{\sigma}_{ab\rightarrow cd}/d\hat{t}$ is the cross section of
parton collisions at leading order, these processes are:
$q\bar{q}\rightarrow q'\bar{q}'$, $qq'\rightarrow qq'$,
$q\bar{q}'\rightarrow q\bar{q}'$, $qq\rightarrow qq$,
$q\bar{q}\rightarrow q\bar{q}$, $qg\rightarrow qg$,
$q\bar{q}\rightarrow gg$, $gg\rightarrow q\bar{q}$ and
$gg\rightarrow gg$ \cite{partco1}. One should note that the gluon
jets contribute only at higher order. $K_{jet}$ is the pQCD
correction factor to take into account the next-to-leading order
(NLO) effects, we use $K_{jet}=$ 1.7 for RHIC and 1.6 for LHC
\cite{Hy3}.

The yield for producing jets in the relativistic heavy ion
collisions is given by
\begin{eqnarray}\label{eq12}
\frac{dN_{jet}}{d^{2}p_{T}dy}=T_{AA}\frac{d\sigma_{jet}}{d^{2}p_{T}dy}(y=0),
\end{eqnarray}
where $T_{AA}=9A^{2}/(8\pi R^{2}_{\bot})$ is the nuclear thickness
for central collisions \cite{Hy1,Hy2}.

\begin{figure}[t]
\begin{center}
\includegraphics[width=9 cm]{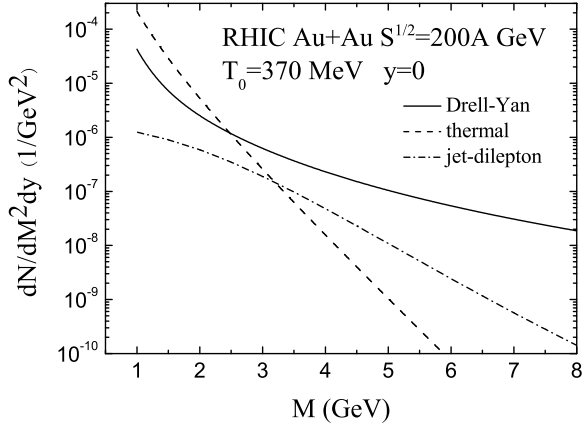}
\vspace{-4ex}\caption{Dilepton yield for central Au+Au collisions at
$\sqrt{s_{NN}}=$200 GeV. We show the dileptons from the QGP (dash
line), dileptons from the Drell-Yan process (solid line), and
dileptons from the passage of jets passing through the hot and dense
plasma (dash dot line).} \label{fig1}
\end{center}
\end{figure}

\begin{figure}[t]
\begin{center}
\includegraphics[width=9 cm]{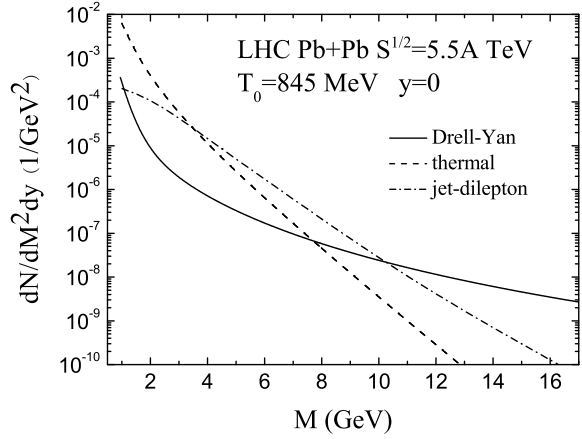}
\vspace{-4ex}\caption{Same as Fig.\ref{fig1} but for central Pb+Pb
collisions at $\sqrt{s_{NN}}=$5.5 TeV.} \label{fig2}
\end{center}
\end{figure}

\subsection{Drell-Yan process}
In the central collisions of two equal-mass nuclei with mass number
$A$ the yield for producing Drell-Yan pairs with the invariant mass
$M$ and rapidity $y$ can be obtained as \cite{jetD1}
\begin{eqnarray}\label{eq13}
\frac{dN_{DY}}{dM^{2}dy}=T_{AA}\frac{d\sigma_{DY}}{dM^{2}dy}(y=0)
\end{eqnarray}
in terms of the cross section of the Drell-Yan process in
nucleon-nucleon collisions \cite{DY1},
\begin{eqnarray}\label{eq13}
\frac{d\sigma_{DY}}{dM^{2}dy}&\!\!\!\!\!\!=\!\!\!\!\!\!&K\frac{4\pi\alpha^{2}}{9M^{4}}\sum_{q}e_{q}^{2}
[x_{a}G_{q/A}(x_{a},Q^{2})x_{b}G_{\bar{q}/B}(x_{b},Q^{2})   \nonumber\\[1mm]
&&+x_{a}G_{\bar{q}/A}(x_{a},Q^{2})x_{b}G_{q/B}(x_{b},Q^{2})],
\end{eqnarray}
where the momentum fractions with rapidity $y$ are
\begin{eqnarray}\label{eq14}
x_{a}=\frac{M}{\sqrt{s_{NN}}}e^{y},
\end{eqnarray}
\begin{eqnarray}\label{eq14}
x_{b}=\frac{M}{\sqrt{s_{NN}}}e^{-y}.
\end{eqnarray}
Here the nuclear effects are considered. A $K$ factor of 1.5 is used
to account for the NLO corrections \cite{NA1}.

\section{Numerical results and discussion}
Figures \ref{fig1} and \ref{fig2} present our results for thermal
dileptons, dileptons from the Drell-Yan process, and dileptons from
the jet-dilepton conversion in the hot and dense plasma at RHIC and
LHC energies, respectively. We find that the contribution of the
jet-dilepton conversion is not prominent at RHIC energies. The
Drell-Yan process is found to dominate over thermal dilepton
emission and the jet-dilepton conversion in the region of $M>$2.5
GeV at RHIC energies(see Fig.\ref{fig1}). The spectrum of dileptons
from the passage of jets interacting with thermal partons falls off
with the invariant mass $M$ faster than the spectrum of the
Drell-Yan process at RHIC. The jet-dilepton conversion starts
playing an interesting role at LHC energies. The jet-dilepton
conversion is the dominant source of large mass dileptons in the
range of 4 GeV$<M<$10 GeV at LHC energies(see Fig.\ref{fig2}). The
spectrum of the jet-dilepton conversion drops rapidly with the $M$
for $M>$10 GeV by comparing with the Drell-Yan process at LHC.

The higher order (NLO) pQCD corrections are accounted for by a
energy and $p_{T}$ dependent $K$ factor, but in the high-energy
collisions the $p_{T}$ dependence of the $K$ factor is weak,
therefore the $K$ factor is assumed to be constant \cite{jetD1,Hy2}.
No $K$ factor for the Drell-Yan process has been used in
Ref.\cite{jetD1}, and the $K_{jet}$ factor of 2.5 used for both RHIC
and LHC in Ref. \cite{jetD1} is larger than the $K_{jet}$ factor
used in this article.

The jets produced in initial parton collisions are defined by all
partons with transverse momentum $p_{T}^{jet}\gg$1 GeV
\cite{jetD1,jetD2}. The dilepton production is sensitive to the
choice of the cutoff $p_{T}^{jet}$. In order to avoid such
sensitivity, the authors of Ref. \cite{jetD1} have constrained a
lower cutoff $p_{T}^{jet}\geq$4 GeV at RHIC and LHC energies. We
adopt this limit in the integration of Eq.(\ref{eq5}).

The main background for the dilepton production in the intermediate
and large mass region is the decay of open charm and bottom mesons.
The $c\bar{c}$($b\bar{b}$) pairs are produced from the initial hard
scattering of partons and can thereafter fragment into $D$($B$) and
$\bar{D}$($\bar{B}$) mesons. If the energy loss of heavy quarks
crossing the hot medium is considered, the contribution of the decay
of open charm and bottom mesons will be suppressed \cite{CEL1,CEL2}.
In this article, this background is not considered, the background
of $J/\Psi$ vector meson decay is also not concerned.

\section{Summary}
We have calculated the production of dileptons produced from the
passage of jets interacting with thermal partons in the hot and
dense plasma. We have rigorously derived the production rate of the
jet-dilepton conversion by using the relativistic kinetic theory.
The spectrum of the jet-dilepton conversion has been compared with
the thermal dilepton emission and the Drell-Yan process. The
numerical results indicate that the contribution of the jet-dilepton
conversion is not prominent for all values of the invariant mass $M$
at RHIC energies. However, this contribution becomes evident at LHC
energies. The jet-dilepton conversion is the dominant source of
large mass dileptons in the range of 4 GeV $<M<$10 GeV at LHC.

\section{Acknowledgements}
This work is supported by the National Natural Science Foundation of
China under Grant Nos 10665003 and 11065010.


%% References


\bibliographystyle{model1-num-names}
\bibliography{<your-bib-database>}

\begin{thebibliography}{99}

%%%%%%%%%%%%%%%%%%%%%%%%%%%%%%%%%%%%%%%%D-Y

\bibitem{DY1}
S. D. Drell, T. M. Yan, Phys. Rev. Lett. {25} (1970) {316}.

%%%%%%%%%%%%%%%%%%%%%%%%%%%%%%%%%%%%%%Thermal

\bibitem{th1}
K. Kajantie, J. Kapusta, L. Mclerran, A. Mekjian, Phys. Rev. D {34}
(1986) {2746}.

\bibitem{th2}
J. Kapusta, L. D. Mclerran, D. K. Srivastava, Phys. Lett. B {283}
(1992) {145}.

\bibitem{th3}
T. Altherr, P. V. Ruuskanen, Nucl. Phys. B {380} (1992) {377}.

\bibitem{th4}
E. V. Shuryak, Phys. Lett. B {78} (1978) {150}.


%%%%%%%%%%%%%%%%%%%%%%%%%%%%%%%%%%%%%%%%%%%%%%%%HG
\bibitem{HG1}
C. Gale, P. Lichard, Phys. Rev. D {49} (1994) {3338}.

\bibitem{HG2}
R. Rapp, J. Wambach, Eur. Phys. J. A {6} (1999) {415}.

\bibitem{HG3}
C. Gale, Nucl. Phys. A {698} (2002) {143}.
%%%%%%%%%%%%%%%%%%%%%%%%%%%%%%%%%%%%%%%%%%%%jet-dilepton
\bibitem{jetD1}
D. K. Srivastava, C. Gale, R. J. Fries, Phys. Rev. C {67} (2003)
{034903}.

\bibitem{jetD2}
S. Turbide, C. Gale, D. K. Srivastava, R. J. Fries, Phys. Rev. C
{74} (2006) {014903}.
%%%%%%%%%%%%%%%%%%%%%%%%%%%%%%%%%%%%%%%%%%%%%%%%%%%%%%%Freze-out
\bibitem{FZ1}
C. M. Hung, E. V. Shuryak, Phys. Rev. C {56} (1997) {453}.

\bibitem{FZ2}
L. G. Landsberg, Phys. Rep. {128} (1985) {301}.

\bibitem{FZ3}
E. V. Shuryak, Rev. Mod. Phys. {65} (1993) {1}.

%%%%%%%%%%%%%%%%%%%%%%%%%%%%%%%%%%%%%%%%%%%%%%%%%%%Dilepton experiments RHIC
\bibitem{DE1}
A. Drees, Nucl. Phys. A {830} (2009) {435}.

\bibitem{DE2}
A. Toia et al., Eur. Phys. J. C {49} (2007) {243}.

\bibitem{DE3}
A. Adare et al., Phys. Rev. Lett. {104} (2010) {132301}.

\bibitem{DE4}
A. Adare et al., Phys. Rev. C {81} (2010) {034911}.

%%%%%%%%%%%%%%%%%%%%%%%%%%%%%%%%%%%%%%%%%%%%%%%%%%%Dilepton experiments SPS
\bibitem{DESPS1}
G. Agakichev et al., Phys. Rev. Lett. {75} (1995) {1272}.
%%%%%%%%%%%%%%%%%%%%%%%%%%%%%%%%%%%%%%%%%%%%%%%%%%%%%%%%%%%%%%%%%chiral symetry
\bibitem{ch1}
R. Rapp, J. Wambach, Adv. Nucl. Phys. {25} (2000) {1}.

%%%%%%%%%%%%%%%%%%%%%%%%%%%%%%%%%%%%%%%%%%%%%%%NA50 NA60
\bibitem{NA1}
R. Rapp, E. Shuryak, Phys. Lett. B {473} (2000) {13}.

\bibitem{NA2}
E. Scomparin et al., Nucl. Phys. A {610} (1996) {331}.

\bibitem{NA3}
M. Floris et al., arXiv:0806.0577 [nucl-ex].
%%%%%%%%%%%%%%%%%%%%%%%%%%%%%%%%%%%%%%%%%%%%Bjorken
\bibitem{Bjor1}
J. D. Bjorken, Phys. Rev. D {27} (1983) {140}.

%%%%%%%%%%%%%%%%%%%%%%%%%%%%%%%%%%%%%%%%%%%%%hydrodynamics and photons
\bibitem{Hy1}
R. J. Fries, B. M$\ddot{\mathrm{u}}$ller, D. K. Srivastava, Phys.
Rev. Lett. {90} (2003) {132301}.

\bibitem{Hy2}
R. J. Fries, B. M$\ddot{\mathrm{u}}$ller, D. K. Srivastava, Phys.
Rev. C {72} (2005) {041902}.

\bibitem{Hy3}
S. Turbide, C. Gale, S. Jeon, G. D. Moore, Phys. Rev. C {72} (2005)
{014906}.

%%%%%%%%%%%%%%%%%%%%%%%%%%%%%%%%%%%%%%%%%%%%%%%%%%%%Energy Loss
\bibitem{EL1}
M. Gyulassy, M. Pl$\ddot{\mathrm{u}}$mer, Phys. Lett. B {243} (1990)
{432}.

\bibitem{EL2}
X. N. Wang, M. Gyulassy, M. Pl$\ddot{\mathrm{u}}$mer, Phys. Rev. D
{51} (1995) {3436}.

\bibitem{EL2.1}
P. Arnold, G. D. Moore, L. Yaffe, J. High Energy Phys. {11} (2001)
{057}.

\bibitem{EL3}
R. Baier et al., Nucl. Phys. B {484} (1997) {265}.
%%%%%%%%%%%%%%%%%%%%%%%%%%%%%%%%%%%%%%%%%%%%%%%%%%%%Owen
\bibitem{Owen}
J. F. Owens, Rev. Mod. Phys. {59} (1987) {465}.

%%%%%%%%%%%%%%%%%%%%%%%%%%%%%%%%%%%%%%%%%%%%nuclear shadowing
\bibitem{shad1}
K. J. Eskola, V. J. Kolhinen, C. A. Salgado, Eur. Phys. J. C {9}
(1999) {61}.

%%%%%%%%%%%%%%%%%%%%%%%%%%%%%%%%%%%%%%%%%%%%%%%%%%%parto distribution
\bibitem{part1}
M. Gl$\ddot{\mathrm{u}}$ck, E. Reya, A. Vogt, Z. Phys. C {67} (1995)
{433}.

%%%%%%%%%%%%%%%%%%%%%%%%%%%%%%%%%%%%%%%%%%%%%%%%%%%%parton colliiosn
\bibitem{partco1}
B. L. Combridge, J. Kripfganz, J. Ranft, Phys. Lett. B {70} (1977)
{234}.

%%%%%%%%%%%%%%%%%%%%%%%%%%%%%%%%%%%%%%%%%%%%%%charm energy loss
\bibitem{CEL1}
G. D. Moore, D. Teaney, Phys. Rev. C {71} (2005) {064904}.

\bibitem{CEL2}
Z. Lin, R. Vogt, X. N. Wang, Phys. Rev. C {57} (1998) {899}.

%%%%%%%%%%%%%%%%%%%%%%%%%%%%%%%%%%%%%%%%%%%%%%%%%%%%%%%%%%%%%%%%%%%%%%%
%\bibitem{q}
%R. D. Field,  \emph{Applications of Perturbative QCD}
%(Addison-Wesley Publishing Company, 1989).

%\bibitem{ww}
%S. Turbide, C. Gale, S. Jeon, G. D. Moore, Phys. Rev. C {72} (2005)
%{014906}.

%\bibitem{ee}
%Z. Kang, J. W. Oiu, W. Vogelsang, Phys. Rev. D {79} (2009) {054007}.

%\bibitem{rr}
%Z. Kang, J. W. Oiu, W. Vogelsang, Nucl. Phys. A {830} (2009) {571}.

%\bibitem{yy}
%E. L. Berger, J. W. Oiu, X. F. Zhang, Phys. Rev. D {65} (2002)
%{034006}.










\end{thebibliography}






% \end{thebibliography}

\end{document}